# Two-Dimensional Scanning Phased Array Based on Pattern Reconfigurable Antenna

Ze Yan[1], Naibo Zhang[2], Guangcun Shan[1,*]
1. School of Instrumentation Science & Opto-electronics Engineering, Beihang university, Beijing, 100191, China.
2. Beijing Research and Development Center, the 54th Research Institute, Electronics Technology Group Corporation, Beijing 100070, China.
*Email: gcshan@buaa.edu.cn

*Abstract*-This paper presents a novel planar phased array that can work in Ka band designed for two-dimensional scanning. A four-modes pattern reconfigurable element based on microstrip Yagi antenna is proposed, and then developed into the 4×4 phased array, in which two feed points are introduced to reduce sidelobe level (SLL). The simulated results show that our proposed phased array is able to cover the scanning range from -60° to +60° in H-plane(*yoz*-plane) and -47.2° to +47.2° in E-plane(*xoz*-plane) with the gain reduction less than 3dB, both the SLLs of which are 10dB less than main beam. The two-dimensional scanning phased array based on pattern reconfigurable antenna (PRA) presented here is suitable for the satellite communication.

## I. INTRODUCTION

Phased arrays have been applicated in many fields due to its high gain and beamforming [1]. With the phase of elements being shifted, antenna arrays can direct their main beam from one angle to another. In modern communication systems, one of the main problems of phased arrays with low side lobes and 3dB main beam gain reduction is the narrow angle scanning range. Therefore, a lot of special antenna elements with features such as wide beam and reconfigurable pattern have been designed to meet the needs of application. However, wide beam antennas generally cause high grating lobes and side lobes which decrease the main beam gain of phased arrays when scan to large angles [2]. On the other hand, pattern reconfigurable antennas (PRAs) are developed for wide-angle scanning phased array [3], [4]. A millimeter-wave phased array based on PRA which formed by three microstrip patches presented in [5], each element can reconfigure their patterns at three different modes with the VSWR less than 2 at 35GHz, and it can scan from -75° to +75° in the elevation plane with a gain fluctuation less than 3dB. Moreover, a genetic algorithm is used to lower the sidelobe levels (SLL). In Ref. [6], a dual-mode PRA is designed for linear phased array. By switching the PRAs between two symmetrical modes, the phased array can cover the scanning range about 162°. The PRAs always reconfigure in two symmetrical directions and need complex bias network. Therefore, they are commonly used in linear phased array instead of a planar array. Recently, a PRA which can steer the main beam in four subspaces has been adopted for a planar phased array [7], and its center frequency is 5.4GHz. At present, PRAs used in planar phased arrays still need further development in simple structures and wide bandwidth, especially in millimeter-wave band.

In this paper, two feed points PRAs are designed for 4×4 wide-angle scanning phased array. By controlling the states of pin diodes, elements can work at right-mode or left-mode. PRAs can also work at front-mode or back-mode by selecting the port which feeds. The structures, features and simulated performances of our proposed PRAs are presented.

## II. PATTERN RECONFIGURABLE ANTENNA DESIGN

The geometry of our designed antenna is shown in Fig. 1. A microstrip Yagi antenna is used which can operate at the frequency of about 28.85GHz. Because of its small size and simple structure, microstrip Yagi antenna has been designed as PRA [8]. Four pin diodes (K1~K4) are installed in gaps, their states determined which strip line plays the role of reflector and another one regarded as a director. Moreover, a finite ground plane is introduced to direct the radiation of the antenna into the upper half space. Due to the SMA probe breaks the symmetry of antenna, two symmetrical feed points (P1, P2) are designed at the center driven strip which is approximately half wavelength at the resonant frequency. Rogers5880 is used as substrate with permittivity of 2.2 and thickness of 1.702mm,

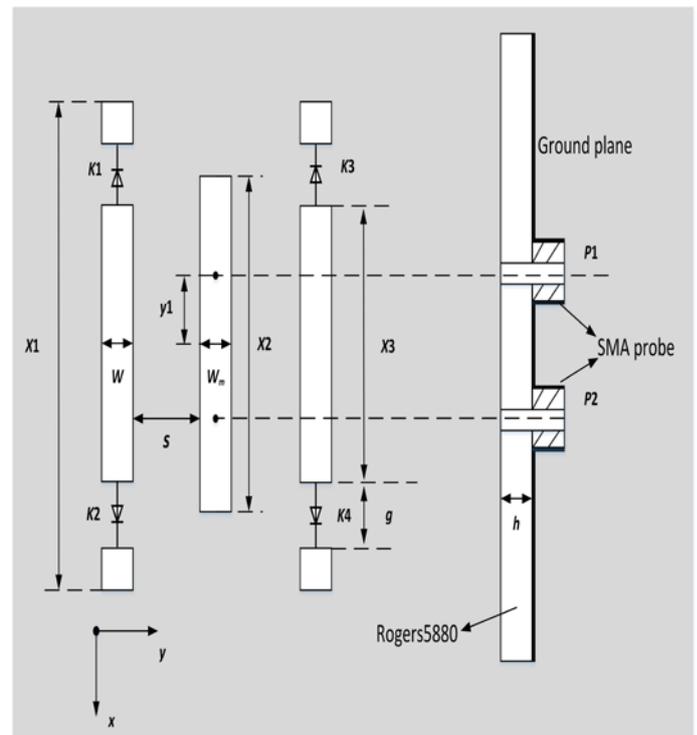

then other detailed parameters shown in Fig. 1 are: $x_1$=4.4mm, $x_2$=4mm, $x_3$=3.05mm, $g$=0.4mm, $s$=1.1mm, $w$=0.2mm,

Figure 1. Structure of the pattern reconfigurable antenna with two feed ports.

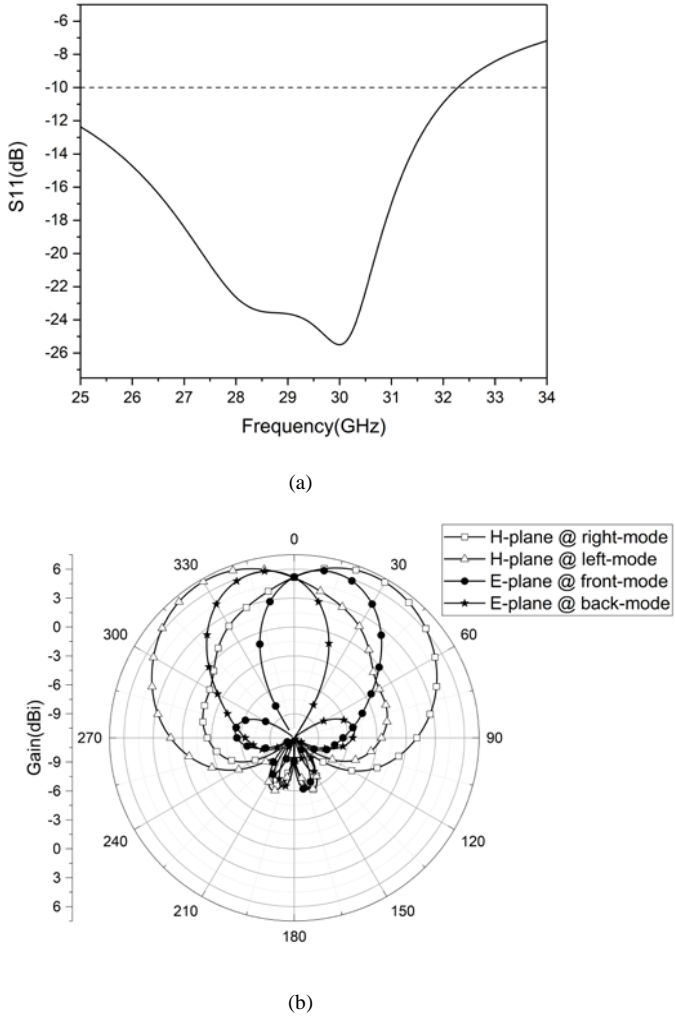

(a)

(b)

Figure 2. (a)Simulated reflection coefficient of antenna and (b)radiation pattern at four modes.

$w_m$=0.8mm, $y_1$=1.2mm. The global dimension is less than 0.5$\lambda$×0.5$\lambda$, which is an advantage for phased arrays.

This antenna can operate in four effective modes by controlling states of switches and ports, specially, only one port works at a time. For example, when K1 and K2 are closed, K3 and K4 are opened, the antenna operates in right-mode regardless of feeding by which port, and the main beam tilts to the positive $y$ axis as shown in Fig. 2(b). In this mode, the left strip is longer than the center strip, with the right is shorter. Therefore, the left element acts as a reflector and the right acts as a director. Conversely, if K1 and K2 are opened, K3 and K4 are closed, the antenna operates in left-mode. On the other hand, when P1 on and P2 off, the antenna operates in back-mode, the radiation pattern in E-plane tilts to the negative $x$ axis more, which is beneficial to lower the SLL of array. Hence, as PRA operating modes switched between back-mode and front-mode, wider scanning angle in E-plane achieved. Fig. 2(a) shows the reflection coefficient (S11) of the antenna. Owing to perfectly symmetric in E-plane and H-plane, the antenna can switch between four modes with the S11 same.

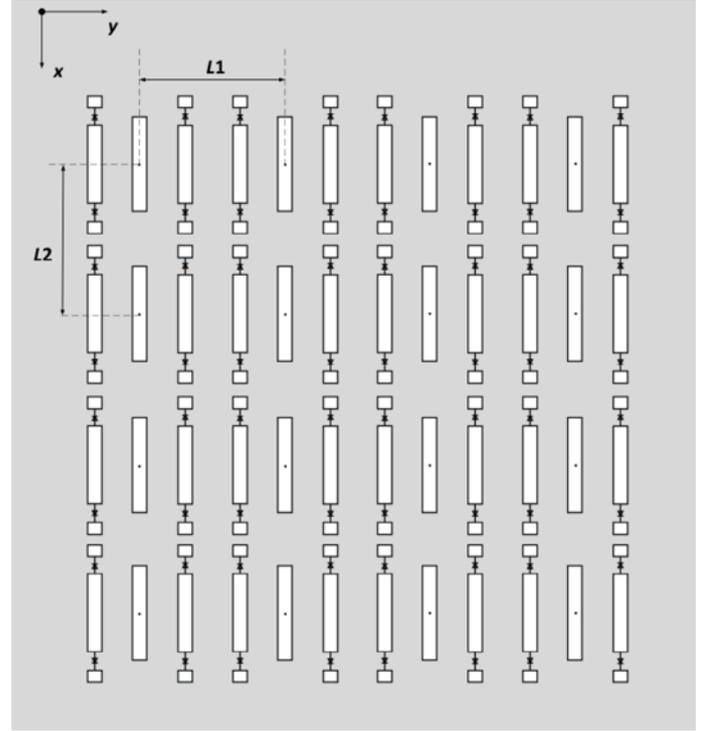

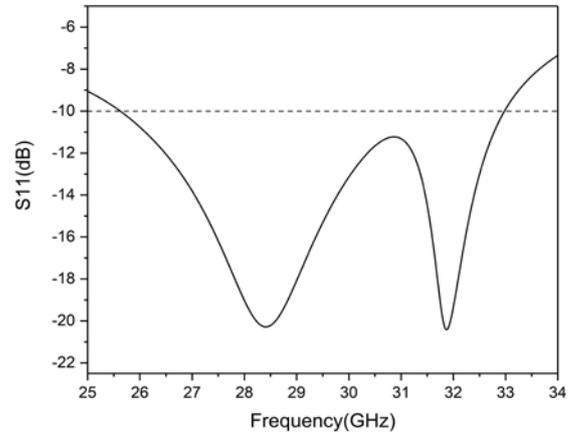

Figure 3. Structure of the proposed phased array based on 16 pattern reconfigurable elements.

Figure 4. Simulated reflection coefficient of the proposed array.

### III. PLANAR PHASED ARRAY WITH PRAs

The above PRAs are used as basic elements of planar phased array. Fig. 3 displays the layout of the overall phased array structure, there are 16 elements in the proposed array. Optimized distance between elements along the $y$ axis and $x$ axis are $L$1=4.5mm, and $L$2=7mm respectively. As shown in Fig. 4, the reflection coefficient of planar phased array is simulated. This result indicates a 25.6% relative bandwidth that S11 less than -10dB, covering from 25.6GHz to 33GHz with center frequency of 28.85GHz. As previously mentioned, the input impedance always changes when antenna switched among reconfigurable modes, making available bandwidth not satisfy the requirement. So, PRA with steady input impedance and wide bandwidth is important to phased array.

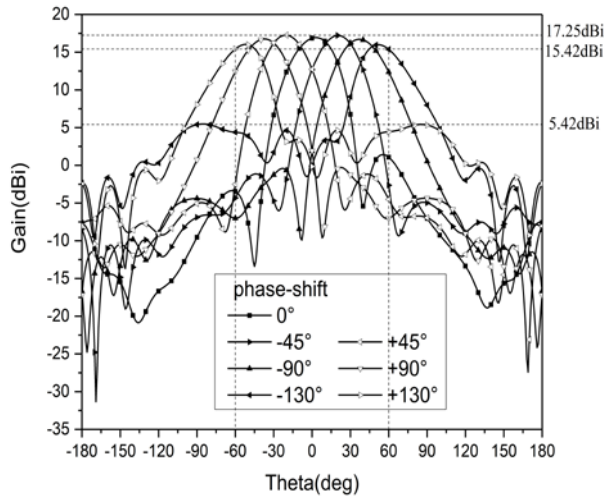

Figure 5. Simulated scanning performance of the proposed array at 28.85GHz in H-plane with a uniformly increasing relative phase-shift among every column, while remains zero among every row.

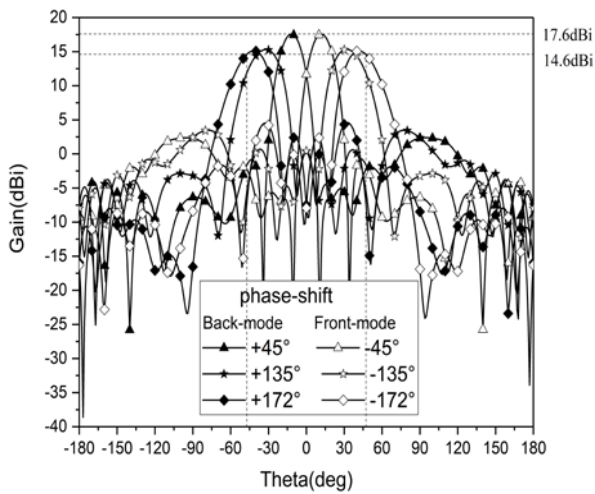

Figure 6. Simulated scanning performance of the proposed array at 28.85GHz in E-plane with a uniformly increasing relative phase-shift among every row, while remains zero among every column.

Fig. 5 exhibits simulated far field radiation pattern in H-plane at the center frequency of array. When phased array worked at right-mode, the radiation pattern tilts to positive $y$ axis by shifting the phases with a lower SLL than left-mode. Inversely, a lower SLL at negative $y$ axis achieved by switching the array to left-mode with the same amount phase shift. By switching pin diodes states and shifting phases of array, scanning range from -60° to +60° in H-plane achieved. The maximum gain is 17.25dBi, and gain at ±60° is 15.24dBi with the -10dB less side lobes in the range of scanning.

As previously mentioned, there is a difference in radiation pattern performed in E-plane to which feed point we choose. Although radiation pattern of element almost remains the same if only one port is designed, there is an obvious scanning performance difference presented between the negative and positive $x$ axis that caused by the asymmetry of element. In order to broaden scanning range in E-plane, two SMA probes are introduced and so two modes are increased. When the elements work at back-mode, SLL decreased in E-plane with main beam of array scanning from 0° to -47.2°. In the same, if scanning range cover from 0° to +47.2°, the state of ports should be switched to front-mode. In this way, we can extend the distance along $x$ axis and thus achieving higher gain. Meanwhile, the increased SSL with longer distance is reduced to satisfy the application requirement. As shown in Fig. 6, a wide 3dB scanning range has also been achieved in the scanning performance in E-plane.

## IV. CONCLUSION

This paper presents a novel planar phased array based on pattern reconfigurable antennas. Microstrip Yagi antenna is used to design PRAs, two symmetrical feed points are designed to decrease SLL in E-plane. Our results of the designed phased array show that array has 25.6% relative bandwidth that is able to scan its main beam from -60° to +60° in H-plane and -47.2° to +47.2° in E-plane with a low SLL. The planar array is designed for Ka band, which could meet the requirement of wide-angle planar scanning for the satellite communication.

ACKNOWLEDGMENT

This work is supported by the National Key R&D Program of China (Grant No.2016YFE0204200).